\documentclass[aps,10pt,prl,notitlepage,tightenlines,nofootinbib,twocolumn,superscriptaddress]{revtex4-2}

\usepackage{amsmath}
\usepackage{amssymb,amsthm}
\usepackage{mathtools}
\usepackage{mathrsfs}
\usepackage{relsize}
\usepackage{bm}
\usepackage{bigstrut}

\usepackage{epsfig,graphics,graphicx,color,xcolor}
\usepackage{soul} 
\usepackage{cancel} 
\usepackage{slashed}	
\usepackage{subfigure}
\usepackage{array}
\usepackage{ragged2e}
\usepackage{lineno}
\usepackage{natbib}
\usepackage{hyperref}
\hypersetup{colorlinks=true,linkcolor=magenta,anchorcolor=red,citecolor=magenta, filecolor=brown,urlcolor=magenta,bookmarksnumbered=true,
pdfview=FitB
}
\graphicspath{ {image/} }

\usepackage{enumitem}
\usepackage{ulem}
\usepackage{soul}

\colorlet{darkgreen}{green!50!black}
\colorlet{brightyellow}{yellow!75!red}
\colorlet{orange}{red!50!yellow}
\colorlet{darkgray}{gray!50!black}
\colorlet{lightblue}{white!50!blue}

\def\dd{{\mathrm{d}}}

\makeatletter
\newcommand*{\transpose}{%
  {\mathpalette\@transpose{}}%
}
\newcommand*{\@transpose}[2]{%
  \raisebox{\depth}{$\m@th#1\intercal$}%
}
\makeatother

\begin{document}

\title{Shedding light on charmonium}

\author{Zhiguo Wang}
\affiliation{Department of Modern Physics, University of Science and Technology of China, Hefei 230026, China}

\author{Meijian~Li}
\email{meijian.li@usc.es}
\affiliation{Instituto Galego de Fisica de Altas Enerxias (IGFAE), Universidade de Santiago de Compostela, E-15782
Galicia, Spain}

\author{Yang~Li}
\affiliation{Department of Modern Physics, University of Science and Technology of China, Hefei 230026, China}

\author{James P. Vary}
\affiliation{Department of Physics and Astronomy, Iowa State University, Ames, Iowa 50011}

\date{\today}

\begin{abstract}
We investigate E1 radiative transitions within charmonium in a relativistic approach based on light-front QCD. In quantum field theory, two sets of processes are pure E1: $\chi_{c0} \to J/\psi \gamma$ ($\psi\to \chi_{c0}\gamma$) and $h_c \to \eta_c\gamma$ ($\eta_c' \to h_c\gamma$), both involving the $P$-wave charmonia. We compute the E1 radiative decay widths as well as the corresponding transition form factors of various processes including those involving $2P$ states. These observables provide an access to the microscopic structures of the $P$-wave charmonium. We show that our parameter-free predictions are in excellent agreement with the experimental measurements as well as lattice simulations whenever available. 
\end{abstract}

 \maketitle
 
\section{Introduction}
The discoveries of charmonium-like states, e.g., $\chi_{c1}(3872)$ and $\chi_{c0}(3915)$, have sparked renewed interests in the charmonium structure \cite{Brambilla:2010cs, Brambilla:2019esw, Chen:2022asf}. The proximity of their masses to the $D\overline D$ threshold leads to the speculation that at least some of them may be meson molecules \cite{Guo:2017jvc}. 
On the other hand, their quantum numbers are consistent with conventional $c\bar c$ quark model and their masses are also in the vicinity of the $2P$ charmonia in various quark model predictions \cite{Godfrey:1985xj}. Furthermore, decay patterns from the quark models can encapsulate various coupled channel effects \cite{Barnes:2003vb, Barnes:2005pb, Swanson:2006st, Barnes:2007xu, Pennington:2007xr, Godfrey:2008nc, Danilkin:2010cc, Ferretti:2013faa, Ortega:2017qmg, Bruschini:2022bsh, Guo:2012tv, Olsen:2014maa, Zhou:2015uva, Yu:2017bsj}. In any case, the investigation of the microscopic structures of charmonium offers new insights into the nature of the strong force, which, after 50 years of QCD, remains one of the biggest puzzles in physics \cite{Gross:2022hyw}.

The radiative transitions provide a clean probe with variable resolutions to the microscopic composition of the system \cite{Henriques:1976jd, Konigsmann:1986wc, Lakhina:2006vg, Eichten:2007qx, Zhao:2013jza, Guo:2014zva, Deng:2016stx, Belle-II:2018jsg, Hoferichter:2020lap, Yao:2021lus, Ganbold:2021nvj, Hong:2022sht}.  Furthermore, these transitions are sensitive to relativistic effects, which underlines some recent discrepancies between NRQCD and the experimental measurements \cite{Babiarz:2019sfa, Feng:2015uha, Feng:2017hlu, Abreu:2022cco}.  For example, the two-photon decay width in NRQCD converges poorly and deviates from the experimental measurements up to $7\sigma$ in NNLO \cite{Feng:2017hlu}. A possible explanation is that charmonium is an intrinsically relativistic system. And the relativistic effects are stronger for the excited states. Therefore, a systematic investigation of the radiative transitions for both the ground-state $P$-wave charmonia and their excitations is required to obtain a complete picture of these charmonium-like states \cite{Chen:2016bpj, Babiarz:2019sfa, Li:2021ejv}.

\begin{figure}
\centering
\includegraphics[width=0.35\textwidth]{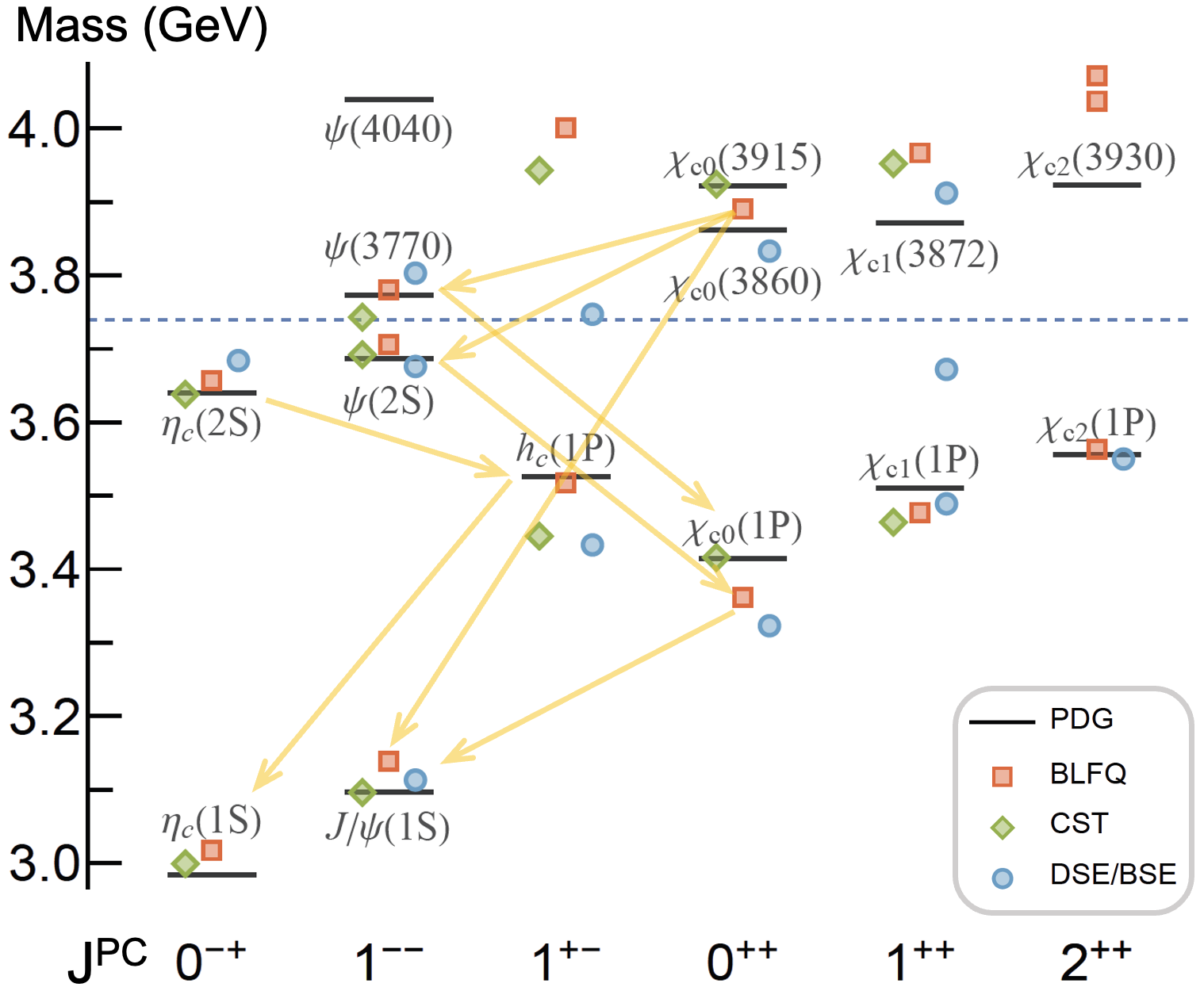}
\caption{Schematic view of the pure E1 transitions (yellow arrows) within charmonium. Other radiative transitions, e.g., M1, M2, and E2, are not shown. The masses obtained in BLFQ along with similar relativistic approaches (CST \cite{Leitao:2017mlx} and DSE/BSE \cite{Fischer:2014cfa}) are shown for comparison (see Sec.~5.4 of Ref.~\cite{Gross:2022hyw} and therein). Figure is adapted from Ref.~\cite{Gross:2022hyw}.}
\label{fig:E1}
\end{figure}

In our previous works, parameter-free predictions are made for the two-photon widths \cite{Li:2021ejv} and the M1 widths \cite{Li:2018uif}, as well as the associated transition form factors (TFFs). 
Our results are based on light-front wave functions (LFWFs \cite{Li:2019data, Maris:2020wew}) from basis light-front quantization (BLFQ \cite{Vary:2009gt}). 
This approach is a natural framework to tackle hadrons as relativistic many-body bound states in the nonperturbative regime \cite{Li:2017mlw}. 
For the application to charmonium, two parameters, the charm quark mass $m_c$ and the basis scale $\kappa$, were fit to the charmonium mass spectrum \cite{Li:2017mlw} (see also Fig.~\ref{fig:E1}). Then the obtained LFWFs are used to make parameter-free predictions to hadronic observables, e.g., decay constants \cite{Li:2017mlw}, as well as partonic observables, e.g., parton distribution functions \cite{Adhikari:2018umb, Chen:2018vdw, Lan:2019img, Babiarz:2023ebe, Lappi:2020ufv}. All of these results were shown to be in reasonable agreement with the experimental measurements whenever available. 

We focus here on the E1 transitions (Fig.~\ref{fig:E1}) between $P$-wave scalar charmonia $\chi_{c0}$  ($0^{++}$) and vector charmonia $\psi$ ($1^{--}$), as well as between $P$-wave axial vector $h_c$ ($1^{+-}$) and pseudoscalars ($0^{-+}$), which are relevant for unraveling the relativistic structure of $P$-wave charmonium. 
We assume the states are pure $c\bar c$'s. Therefore, any significant deviation from the experimental measurements implies a deviation from the conventional $c\bar c$ picture.

\begin{figure}
\centering
\includegraphics[width=0.4\textwidth]{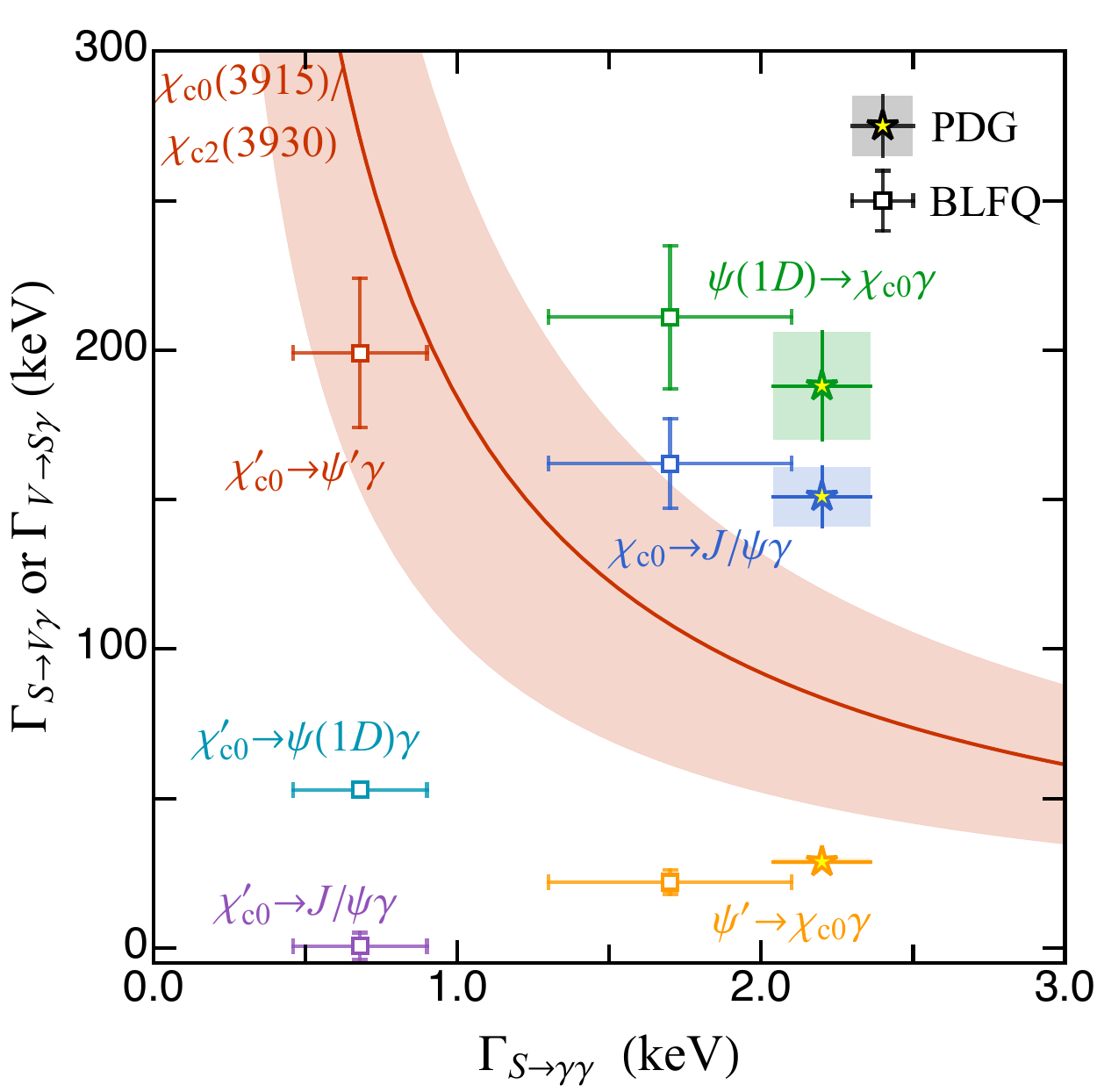}
\caption{(Colors online) The BLFQ prediction of scalar charmonium radiative widths $\Gamma_{S\to\gamma\gamma}$ and $\Gamma_{S\to V\gamma}$ as compared with the PDG values \cite{ParticleDataGroup:2022pth}. The red curve with a band is the Belle measurement of the product $\Gamma(R \to \psi' \gamma) \Gamma(R\to\gamma\gamma)/\Gamma_\text{total} = 9.8(3.6)(1.3) \,\mathrm{eV}$, where $R$ is identified as $\chi_{c0}(3915)$ or $\chi_{c2}(3930)$ \cite{Belle:2021nuv}. }
\label{fig:SVV_widths_combined}
\end{figure}

The E1 transition has a similar helicity structure SVV with the scalar meson two-photon transition \cite{Li:2021ejv}. Since the structures of the photon and vector charmonia are well established, these two processes can be used to constrain the structure of the scalar charmonia at different scales.  
Fig.~\ref{fig:SVV_widths_combined} combines the E1 widths $\Gamma_{S\to V\gamma}$ (or $\Gamma_{V\to S\gamma}$) and the two-photon width $\Gamma_{S\to\gamma\gamma}$ for scalar charmonia $\chi_{c0}(1P)$ and $\chi_{c0}(2P)$, as obtained in BLFQ \cite{Li:2021ejv}. Processes for $1P$ scalar $\chi_{c0}$ have been measured by several experiments and compiled by PDG \cite{ParticleDataGroup:2022pth}. Our results are in good agreement with the PDG values, which provide a basis for making predictions for the $2P$ state $\chi'_{c0}$. 
Experimentally, the only available E1 data come from Belle for $\chi_{c0}(3915)$, a prime candidate for the $2P$ scalar. 
Last year, Belle collaboration discovered a resonance with the mass 3.922 GeV, which can be identified as $\chi_{c0}(3915)$ or $\chi_{c2}(3930)$. Belle also measured the product $\Gamma(R \to \psi' \gamma) \Gamma(R\to\gamma\gamma)/\Gamma_\text{total} = 9.8(3.6)(1.3) \,\mathrm{eV}$, which is shown as a red curve with a band in Fig.~\ref{fig:SVV_widths_combined} \cite{Belle:2021nuv}. Our predicted E1 and diphoton widths are consistent with this result.

\section{Formalism}\label{sec:formalism}

The E1 amplitude of a scalar meson $S$ decaying into a vector meson $V$ plus a (virtual) photon is described by the hadronic matrix element (HME), 
\begin{equation}
H_{\lambda_\gamma \lambda'}(q^2) = e\mathcal Q_c \varepsilon^{\mu*}_{\lambda_\gamma}(q) \langle V(p', \lambda')|J _\mu(0)|S(p)\rangle\;,
\end{equation}
where, $J_\mu(x)$ is the current operator, $\mathcal Q_c=2/3$ is the charge number, $e = \sqrt{4\pi\alpha_\text{em}}$ is the electron charge, $q = p'-p$ is the four-momentum of the photon, and $\varepsilon^{\mu}_{\lambda_\gamma}(q)$ is the polarization vector of the photon. 
Following Ref.~\cite{Dudek:2006ej}, we parametrize the HME in terms of its Lorentz structures, 
\begin{multline} \label{eqn:HME}
\langle V(p', \lambda')|J^\mu(0)|S(p)\rangle = E_1(Q^2) 
 \Big[e^{\mu*}_{\lambda'}(p') - \\ \frac{e_{\lambda'}^* \cdot p}{\Omega(Q^2)}\big(p'^\mu (p\cdot p') - M^2_V p^\mu\big)\Big]  
+ \frac{M_VC_1(Q^2) }{iQ\Omega(Q^2)}(e^*_{\lambda'}\cdot p) \\ 
\times \Big[(p\cdot p')(p+p')^\mu - M_S^2 p'^\mu - M_V^2 p^\mu\Big]\;,
\end{multline}
where, $Q^2 = -q^2$, and $\Omega(Q^2) = (p\cdot p')^2 - M_S^2M_V^2$. $e^\mu_\lambda(p)$ is the polarization vector of the vector meson. The form factors $E_1$ and $C_1$ defined here can be extracted from the transverse and longitudinal amplitudes, respectively, viz.,
\begin{equation}
\begin{split}
H_{\lambda_\gamma=\pm 1, \lambda'} = \,& e\mathcal Q_c E_1(Q^2), \\
H_{\lambda_\gamma=0, \lambda'} =\,& -e\mathcal Q_c C_1(Q^2).
\end{split}
\end{equation}
In particular, the E1 radiative decay width is proportional to $|E_1(0)|^2$, 
\begin{equation}
\Gamma = \frac{\mathcal Q_c^2\alpha_\text{em}}{2j_\text{i}+1} \frac{ M_\text{i}^2-M_\text{f}^2}{2M_\text{i}^3} \big|E_1(0)\big|^2.
\end{equation}
Here, $j_\text{i}, M_\text{i}$ are the initial state spin and mass, respectively, and $M_\text{f}$ is the final state mass. 
The vector meson decaying into a scalar plus a photon can be similarly expressed. The Lorentz structures of the HME between pseudoscalar $0^{-+}$ and the $C$-odd axial vector $1^{+-}$ are identical to Eq.~(\ref{eqn:HME}).

\begin{figure}
\centering
\includegraphics[width=0.4\textwidth]{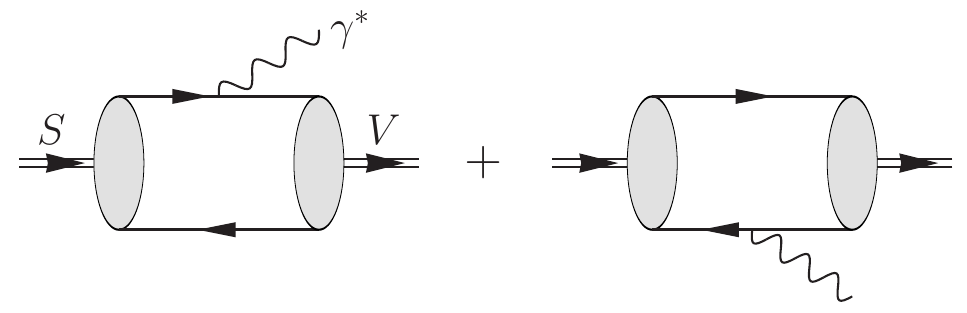}
\caption{Leading order diagrams of the E1 radiative transition $S \to V + \gamma^*$.}
\label{fig:E1_diagrams}
\end{figure}

In a nonrelativistic quark model, the E1 transition is induced by the electric dipole interaction. The corresponding electric charge density on the light front is $J^+ = J^0 + J^3$, where we adopt the light-front coordinates $v^\mu = (v^+, v^-, \vec v_\perp)$ with $v^\pm = v^0 \pm v^3$ and $\vec v_\perp = (v^1, v^2)$. $J^+$ is also known as the ``good current'' in light-front dynamics as it is not contaminated by the spurious Lorentz-symmetry violating contributions \cite{Drell:1969km, Brodsky:1997de, Carbonell:1998rj}. We further adopt the Drell-Yan frame $q^+ = 0$ which simplifies the expression dramatically \cite{Drell:1969km, Carbonell:1998rj}. The relevant diagrams for the E1 process are shown in Fig.~\ref{fig:E1_diagrams}. The nonperturbative structures of the initial- and final-state mesons are encoded in the LFWFs, 
\begin{multline}
|\psi_h(P, j, m_j)\rangle = \sum_{s, \bar s}\int_0^1 \frac{\dd x}{2x(1-x)} \int \frac{\dd^2 k_\perp}{(2\pi)^3} \psi_{s\bar s/h}^{(m_j)}(x, \vec k_\perp) \\
\times \frac{1}{\sqrt{N_c}} \sum_i b^\dagger_{si}(p)  d^\dagger_{\bar si}(\bar p) |0\rangle + \cdots\;,
\end{multline}
where, $x=p^+/P^+$ is the longitudinal momentum fraction of the quark, and $\vec k_\perp = \vec p_\perp -x\vec P_\perp$ is the relative transverse momentum of the quark. The momenta of the quark and the antiquark are $p^\mu = (xP^+, \vec k_\perp+x\vec P_\perp)$, $\bar p^\mu = \big((1-x)P^+, -\vec k_\perp+(1-x)\vec P_\perp\big)$. Here, $i = 1,2,\cdots N_c$ is the color index and $N_c = 3$. The LFWF $\psi_{s\bar s/h}^{(m_j)}(x, \vec k_\perp)$ is frame-independent and only depends on the relative motion of the quark and antiquark. 
The ellipsis represents contributions beyond the valence Fock sector $|c\bar c\rangle$, which are shown to be small from previous investigations and will be neglected for the present work as well (cf.~\cite{Lan:2021wok}).  

Using the LFWFs, the TFF E1 can be represented as
\begin{multline}
E_1(Q^2) = 4\sum_{s, \bar s}\int \frac{\dd x}{2x(1-x)}\int \frac{\dd^2k_\perp}{(2\pi)^3}  \Big\{ M_V\psi_{s\bar s/V}^{(\lambda=0)*}(x, \vec k_\perp)  \\
+  \frac{M_S^2-M_V^2+Q^2}{\sqrt{2}Q} \psi_{s\bar s/V}^{(\lambda=+1)*}(x, \vec k_\perp)
\Big\} 
 \psi_{s\bar s/S}\big(x, \vec k_\perp+(1-x)\vec q_\perp\big)
 \;,
\end{multline}
where, $Q^2 = -q^2 = q_\perp^2$, and we have adopted $\arg \vec q_\perp = 0$ for simplicity. 
The coupling constant $E_1(0)$ is associated with dipole transition between the transversely polarized vector meson and the scalar meson:
\begin{multline}\label{eqn:E1_coupling}
E_1(0) =  \frac{M_S^2-M_V^2}{i\sqrt{2}}  \sum_{s, \bar s}\int_0^1 \frac{\dd x}{4\pi} \int \dd^2 r_\perp  (r_x+ir_y)   \\
 \times \psi_{s\bar s/V}^{(\lambda=+1)*}(x, \vec r_\perp)\psi_{s\bar s/S}(x, \vec r_\perp). 
\end{multline}
This expression resembles the nonrelativistic expression of the E1 transition. 
The TFF can also be extracted from the spatial current $\vec J_\perp$ \cite{Li:2018uif, Li:2021ejv}. 
Since the E1 transition is induced by the electric dipole, we adopt the charge density operator $J^+$, which has a smooth nonrelativistic limit as shown by Eq.~\eqref{eqn:E1_coupling}. 

\section{Numerical results}\label{sec:results}

\begin{figure}
 \subfigure[\ ${\chi_{c0} \to J/\psi + \gamma}$ \label{fig:chic0_to_Jpsi_width}]{\includegraphics[width=0.375\textwidth]{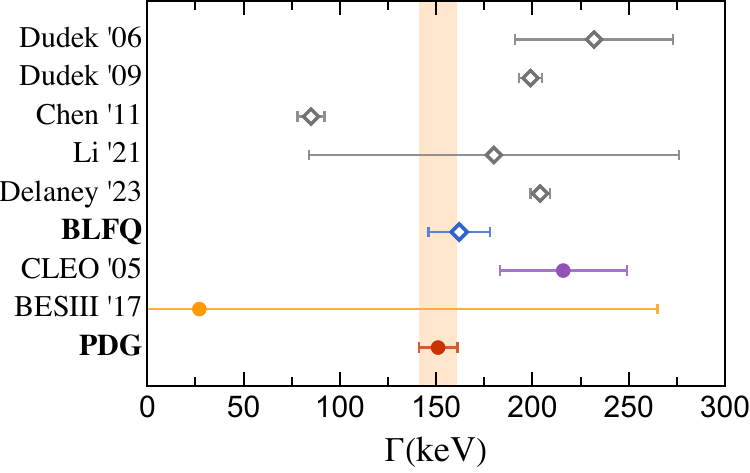}} \\
 \subfigure[\ ${ \psi(2S) \to \chi_{c0} + \gamma}$ \label{fig:chic0_to_psi2S_width}]{\includegraphics[width=0.375\textwidth]{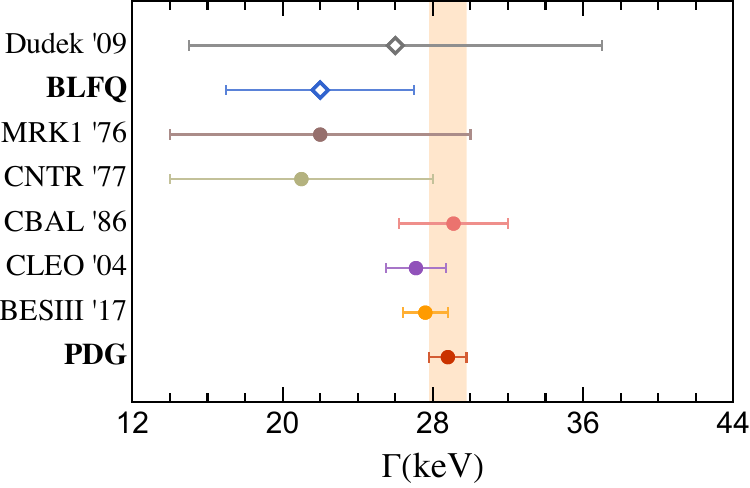}} \\
 \subfigure[\ ${\psi(3770) \to \chi_{c0} + \gamma}$ \label{fig:chic0_to_psi1D_width}]{\includegraphics[width=0.375\textwidth]{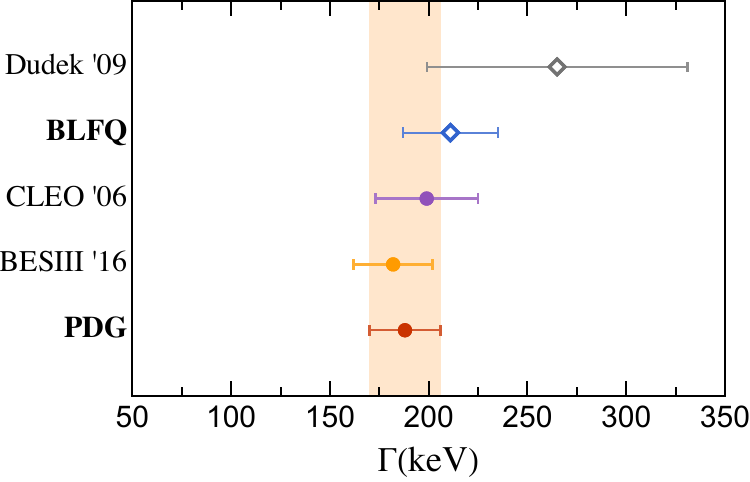}} \\
 \subfigure[\ ${h_c \to \eta_c(1S)+\gamma}$ \label{fig:hc_to_etac1S_width}]{\includegraphics[width=0.375\textwidth]{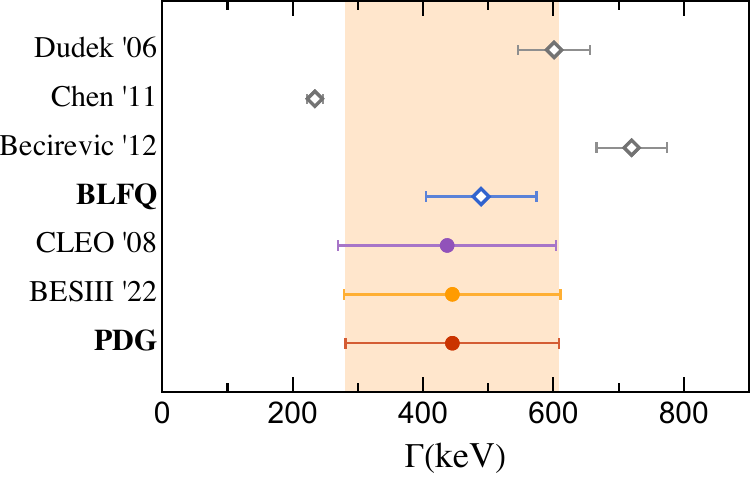}} \hspace*{1em}
\caption{Comparison of the E1 decay widths from this work (BLFQ) and from the experimental measurements \cite{CLEO:2005efp, BESIII:2017gcu,Whitaker:1976hb, Biddick:1977sv, Gaiser:1985ix,CLEO:2004cbu,CLEO:2006nor, BESIII:2015cby, BESIII:2012urf, BESIII:2022tfo, CLEO:2008ero, BESIII:2010gid, CLEO:2005vqq} including the PDG values \cite{ParticleDataGroup:2022pth}. The lattice QCD results \cite{Dudek:2006ej,Dudek:2009kk,Chen:2011kpa,Becirevic:2012dc,Li:2021gze,Delaney:2023fsc} are also shown for comparison. }
\label{fig:E1_width}
\end{figure}

The E1 widths associated with the ground-state scalar $\chi_{c0}$, i.e., $\Gamma_{\chi_{c0} \to J/\psi \gamma}$, $\Gamma_{ \psi(2S) \to \chi_{c0} \gamma}$, and $\Gamma_{\psi(3770) \to \chi_{c0} \gamma}$, and with the $C$-odd axial vector $h_c$, i.e., $\Gamma_{{h_c} \to \eta_c(1S) \gamma}$, $\Gamma_{\eta_c(2S) \to h_c \gamma}$, have been measured by several experiments \cite{CLEO:2005efp,BESIII:2017gcu,Whitaker:1976hb, Biddick:1977sv, Gaiser:1985ix,CLEO:2004cbu, CLEO:2006nor, BESIII:2015cby, BESIII:2012urf, BESIII:2022tfo, CLEO:2008ero, BESIII:2010gid, CLEO:2005vqq} and compiled by PDG \cite{ParticleDataGroup:2022pth}. Some of the most recent measurements come from the CLEO and BES III collaborations. We compare our results\footnote{ The values of the decay widths are provided in the supplementary material of this paper.}   with the experimental data as well as the PDG values in Figs.~\ref{fig:chic0_to_Jpsi_width}--\ref{fig:hc_to_etac1S_width}. Lattice QCD results \cite{Dudek:2006ej,Dudek:2009kk,Chen:2011kpa,Becirevic:2012dc,Li:2021gze,Delaney:2023fsc} are also compared in the same plots.
Overall, our parameter-free results are in excellent agreement with the experimental data.  Following our previous analyses for dilepton and diphoton and radiative transitions \cite{Li:2017mlw, Li:2018uif, Li:2021ejv}, we use the $N_{\text{max}} = 8$ BLFQ LFWFs, whose UV scale $\Lambda_\textsc{uv} = \kappa \sqrt{N_\text{max}} = 2.8\,\mathrm{GeV}$ matches the charmonium scale. We estimate the model uncertainty as the difference between the $N_{\text{max}} = 8$ and $N_\text{max} = 16$ results. 

\begin{figure*}
 \subfigure[\ ${\chi_{c0} \to J/\psi + \gamma}$ \label{fig:chic0_to_Jpsi_TFF}]{\includegraphics[width=0.32\textwidth]{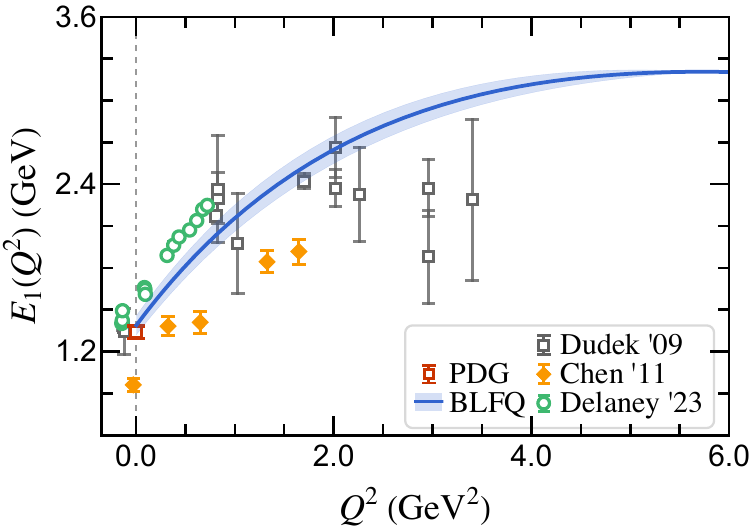}}
 \subfigure[\ ${ \psi(2S) \to \chi_{c0} + \gamma}$ \label{fig:chic0_to_psi2S_TFF}]{\includegraphics[width=0.32\textwidth]{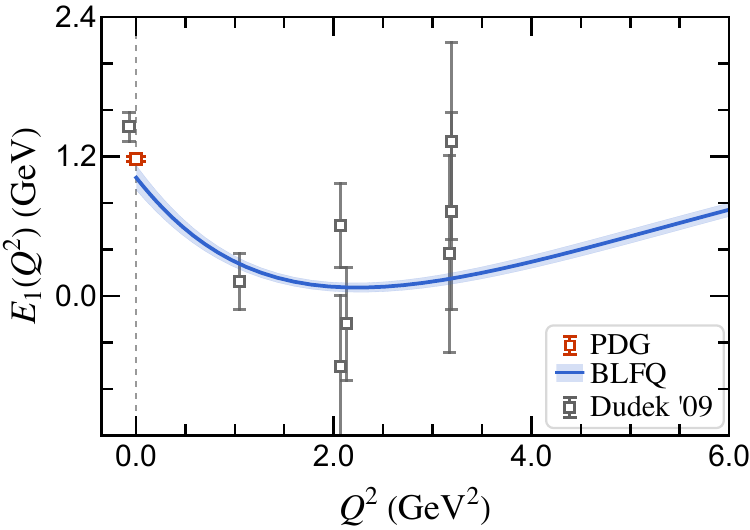}}
 \subfigure[\ ${\psi(3770) \to \chi_{c0} + \gamma}$ \label{fig:chic0_to_psi1D_TFF}]{\includegraphics[width=0.32\textwidth]{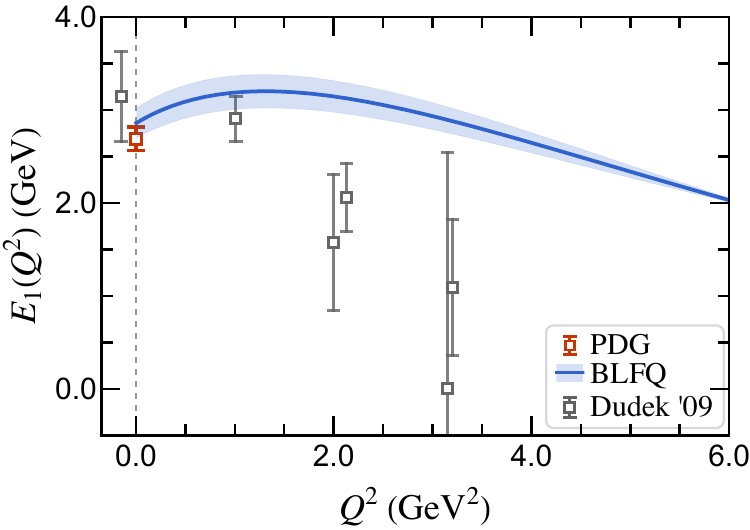}} \\
  \subfigure[\ ${\chi_{c0}(2P) \to J/\psi + \gamma}$ \label{fig:chic02P_to_Jpsi_TFF}]{\includegraphics[width=0.32\textwidth]{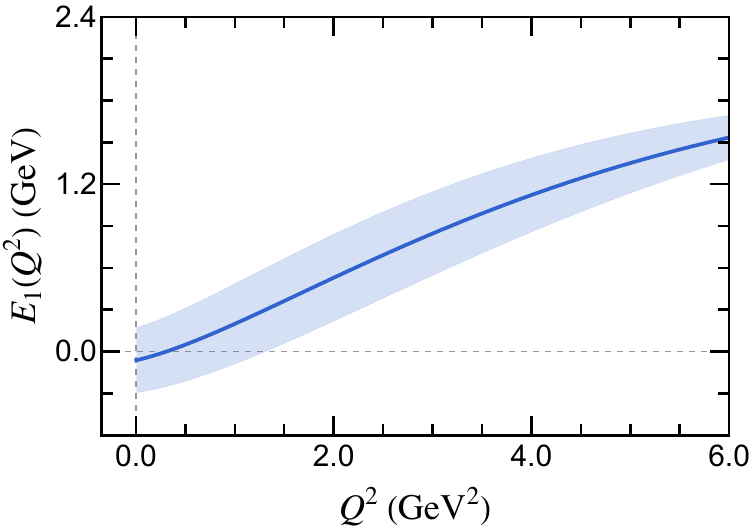}}
 \subfigure[\ ${ \chi_{c0}(2P) \to \psi(2S) +  \gamma}$ \label{fig:chic02P_to_psi2S_TFF}]{\includegraphics[width=0.32\textwidth]{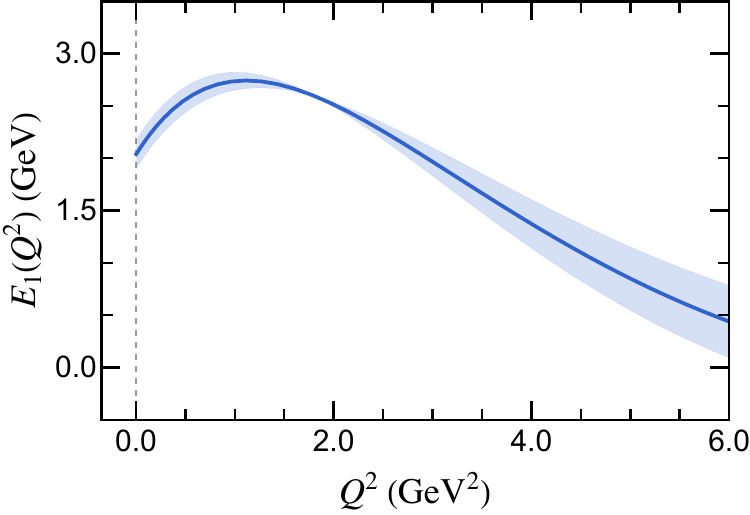}}
 \subfigure[\ ${\chi_{c0}(2P) \to \psi(3770) + \gamma}$ \label{fig:chic02P_to_psi1D_TFF}]{\includegraphics[width=0.32\textwidth]{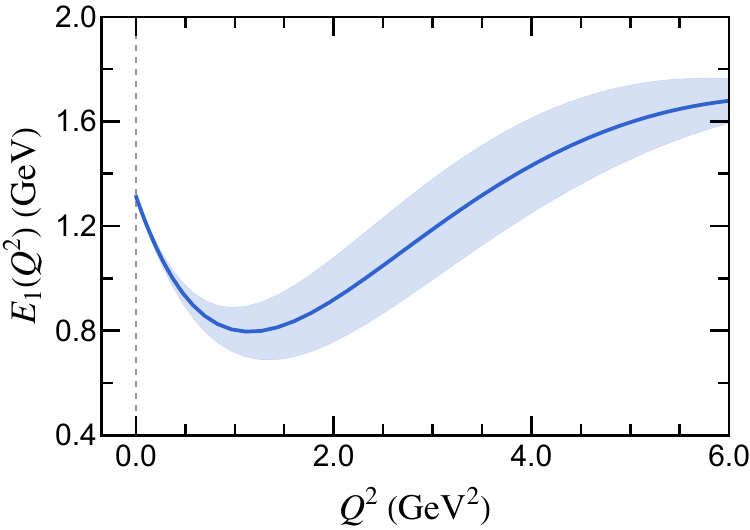}} \\
 \,\, \subfigure[\ ${h_c(1P) \to \eta_c(1S) + \gamma}$ \label{fig:hc_to_etac1S_TFF}]{\includegraphics[width=0.325\textwidth]{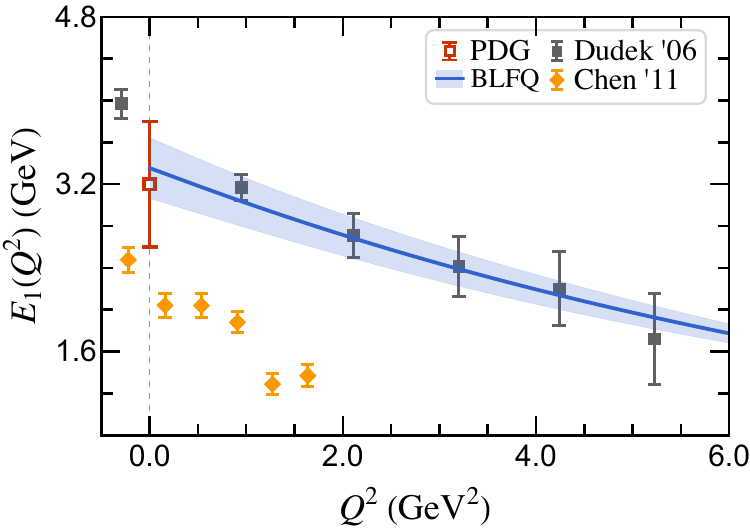}}
   \subfigure[\ ${ \eta_c(2S) \to h_c(1P) + \gamma}$ \label{fig:hc_to_etac2S_TFF}]{\includegraphics[width=0.325\textwidth]{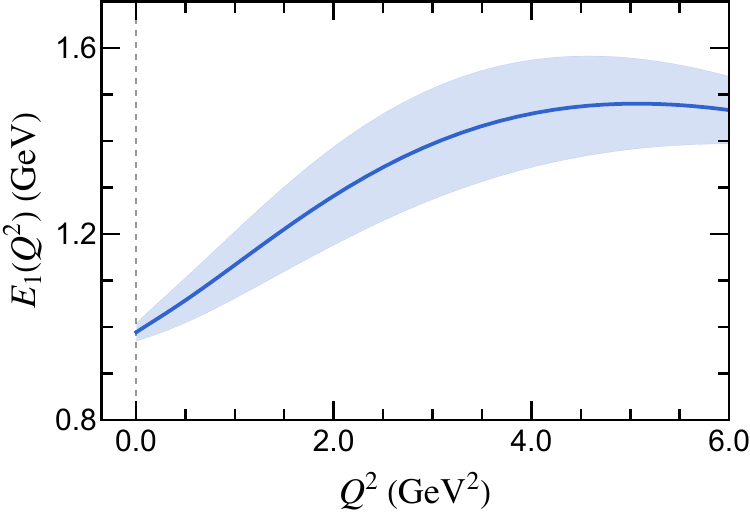}} \hfill { }
\caption{Comparison of the E1 radiative transition form factors from this work (BLFQ) and from several lattice simulations \cite{Dudek:2006ej,Dudek:2009kk,Chen:2011kpa,Delaney:2023fsc}. }
\label{fig:E1_TFF}
\end{figure*}

The TFF $E_1(Q^2)$ provides further resolution of the system. Alas, the TFFs of these processes are not currently available from the experiments. We thus compare our BLFQ results with recent lattice simulations \cite{Dudek:2006ej,Dudek:2009kk,Chen:2011kpa,Delaney:2023fsc}. 
Fig.~\ref{fig:chic0_to_Jpsi_TFF} compares the TFFs of the transition between $\chi_{c0}$ and $J/\psi$ as predicted by BLFQ and several lattice calculations \cite{Dudek:2009kk,Chen:2011kpa,Delaney:2023fsc}. 
Given the scattering of the lattice data from different groups, our BLFQ prediction is in reasonable agreement with these results, in particular at low $Q^2$. Our approach also provides access to moderately high $Q^2$, where the lattice simulations suffer from low statistics. 

Similarly, Figs.~\ref{fig:chic0_to_psi2S_TFF} and \ref{fig:chic0_to_psi1D_TFF} show the E1 TFFs of the processes $\psi(2S) \to \chi_{c0} + \gamma$ and ${\psi(3770) \to \chi_{c0} + \gamma}$, respectively. 
In nonrelativistic pictures, radiative transitions involving radially or angularly excited states,  e.g. $\psi(2S)$ and $\psi(1D)$, are sensitive to the shape of the mesons wave function, e.g., the locations of the nodes. 
 Our BLFQ predictions are again in good agreement with the lattice results \cite{Dudek:2009kk}, albeit the statistics of latter is limited at $Q$ above 1 GeV. 
 Figs.~\ref{fig:chic02P_to_Jpsi_TFF}--\ref{fig:chic02P_to_psi1D_TFF} show the E1 TFFs of $\chi_{c0}(3915)$ to the low-lying vector mesons, assuming that $\chi_{c0}(3915)$ is the $2P$ $c\bar c$ state. These predictions may serve as a benchmark for future investigation of $\chi_{c0}(3915)$. 
 
 The transitions involving the $C$-odd axial vector $h_c$ are shown in Figs.~\ref{fig:hc_to_etac1S_TFF}--\ref{fig:hc_to_etac2S_TFF}. The TFF of the process $h_c \to \eta_c(1S)+\gamma$ is computed by Refs.~\cite{Dudek:2006ej,Chen:2011kpa} in lattice. Our results are in good agreement with Ref.~\cite{Dudek:2006ej} while deviating from \cite{Chen:2011kpa}. Note that 
our E1 width of this process is in better agreement with the experimental data \cite{BESIII:2012urf, BESIII:2022tfo, CLEO:2008ero, BESIII:2010gid, CLEO:2005vqq}.

\section{Summary}\label{sec:summary}

In this work, we investigate the E1 radiative transitions within the charmonium system using the basis light-front quantization approach. We derived the light-front wave function representation of the decay width as well as the transition form factors. These representations are exact as long as the wave functions are exactly known. The wave functions adopted in this work come from fitting to the charmonium spectrum. Therefore, we are able to make parameter-free predictions for the E1 transitions. The results, including the widths and the form factors, are in excellent agreement with the experimental measurements as well as lattice simulations whenever available. 

We also compute the E1 widths and the corresponding transition form factors of $\chi_{c0}(3915)$ by treating it as the $2P$ $c\bar c$ state. The obtained results are consistent with the recent measurement from Belle \cite{Belle:2021nuv}. Further experimental measurements are required to discern the nature of this particle. 

We note similar successes in describing the charmonium structures, viz., M1 transitions \cite{Li:2018uif}, two-photon transitions \cite{Li:2021ejv} as well as the decay constants \cite{Li:2017mlw} using the same set of light front wave functions. These applications provide confidence that the intrinsic structure of charmonium is accurately described by these BLFQ wave functions and lend support to the adopted phenomenological form of confinement \cite{Brodsky:2014yha}. We envision that further applications of the charmonium LFWFs will help to resolve the non-perturbative dynamics of the strong interaction in high-energy processes, such as the gluon distributions, generalized parton distributions (GPDs), hadronic anomalous energy, etc., which are among the central goals of the forthcoming electron-ion colliders \cite{Cisek:2014ala, Chen:2016dlk, Goncalves:2017wgg, Babiarz:2023cac, Aschenauer:2017jsk, Goncalves:2020ywm, Anderle:2021wcy, AbdulKhalek:2021gbh, Mantysaari:2022kdm, Abir:2023fpo, Accardi:2023chb}.

\section{Acknowledgements}

The authors acknowledge valuable discussions with P. Maris and X. Zhao. 

Y.L. is supported by the New faculty start-up fund of the University of Science and Technology of China. 
M.L. is supported by Xunta de Galicia (CIGUS accreditation), European Union ERDF, the Spanish Research State Agency under project PID2020-119632GB-I00, and European Research Council under project ERC-2018-ADG-835105 YoctoLHC. 

 This work was supported in part by the Chinese Academy of Sciences under Grant No.~YSBR-101, and in part by the US Department of Energy (DOE) under Grant No. DE-SC0023692.


\begin{thebibliography}{99}


\bibitem{Brambilla:2010cs}
N.~Brambilla, S.~Eidelman, B.~K.~Heltsley, R.~Vogt, G.~T.~Bodwin, E.~Eichten, A.~D.~Frawley, A.~B.~Meyer, R.~E.~Mitchell and V.~Papadimitriou, \textit{et al.}
Eur. Phys. J. C \textbf{71}, 1534 (2011)
doi:10.1140/epjc/s10052-010-1534-9
[arXiv:1010.5827 [hep-ph]].

\bibitem{Brambilla:2019esw}
N.~Brambilla, S.~Eidelman, C.~Hanhart, A.~Nefediev, C.~P.~Shen, C.~E.~Thomas, A.~Vairo and C.~Z.~Yuan,
Phys. Rept. \textbf{873}, 1-154 (2020)
doi:10.1016/j.physrep.2020.05.001
[arXiv:1907.07583 [hep-ex]].

\bibitem{Chen:2022asf}
H.~X.~Chen, W.~Chen, X.~Liu, Y.~R.~Liu and S.~L.~Zhu,
Rept. Prog. Phys. \textbf{86}, no.2, 026201 (2023)
doi:10.1088/1361-6633/aca3b6
[arXiv:2204.02649 [hep-ph]].

\bibitem{Guo:2017jvc}
F.~K.~Guo, C.~Hanhart, U.~G.~Mei\ss{}ner, Q.~Wang, Q.~Zhao and B.~S.~Zou,
Rev. Mod. Phys. \textbf{90}, no.1, 015004 (2018)
[erratum: Rev. Mod. Phys. \textbf{94}, no.2, 029901 (2022)]
doi:10.1103/RevModPhys.90.015004
[arXiv:1705.00141 [hep-ph]].

\bibitem{Godfrey:1985xj}
S.~Godfrey and N.~Isgur,
Phys. Rev. D \textbf{32}, 189-231 (1985)
doi:10.1103/PhysRevD.32.189

\bibitem{Barnes:2003vb}
T.~Barnes and S.~Godfrey,
Phys. Rev. D \textbf{69}, 054008 (2004)
doi:10.1103/PhysRevD.69.054008
[arXiv:hep-ph/0311162 [hep-ph]].

\bibitem{Barnes:2005pb}
T.~Barnes, S.~Godfrey and E.~S.~Swanson,
Phys. Rev. D \textbf{72}, 054026 (2005)
doi:10.1103/PhysRevD.72.054026
[arXiv:hep-ph/0505002 [hep-ph]].

\bibitem{Swanson:2006st}
E.~S.~Swanson,
Phys. Rept. \textbf{429}, 243-305 (2006)
doi:10.1016/j.physrep.2006.04.003
[arXiv:hep-ph/0601110 [hep-ph]].

\bibitem{Barnes:2007xu}
T.~Barnes and E.~S.~Swanson,
Phys. Rev. C \textbf{77}, 055206 (2008)
doi:10.1103/PhysRevC.77.055206
[arXiv:0711.2080 [hep-ph]].

\bibitem{Pennington:2007xr}
M.~R.~Pennington and D.~J.~Wilson,
Phys. Rev. D \textbf{76}, 077502 (2007)
doi:10.1103/PhysRevD.76.077502
[arXiv:0704.3384 [hep-ph]].

\bibitem{Godfrey:2008nc}
S.~Godfrey and S.~L.~Olsen,
Ann. Rev. Nucl. Part. Sci. \textbf{58}, 51-73 (2008)
doi:10.1146/annurev.nucl.58.110707.171145
[arXiv:0801.3867 [hep-ph]].

\bibitem{Danilkin:2010cc}
I.~V.~Danilkin and Y.~A.~Simonov,
Phys. Rev. Lett. \textbf{105}, 102002 (2010)
doi:10.1103/PhysRevLett.105.102002
[arXiv:1006.0211 [hep-ph]].

\bibitem{Ferretti:2013faa}
J.~Ferretti, G.~Galat\`a and E.~Santopinto,
Phys. Rev. C \textbf{88}, no.1, 015207 (2013)
doi:10.1103/PhysRevC.88.015207
[arXiv:1302.6857 [hep-ph]].

\bibitem{Ortega:2017qmg}
P.~G.~Ortega, J.~Segovia, D.~R.~Entem and F.~Fern\'andez,
Phys. Lett. B \textbf{778}, 1-5 (2018)
doi:10.1016/j.physletb.2018.01.005
[arXiv:1706.02639 [hep-ph]].

\bibitem{Bruschini:2022bsh}
R.~Bruschini and P.~Gonz\'alez,
JHEP \textbf{02}, 216 (2023)
doi:10.1007/JHEP02(2023)216
[arXiv:2207.02740 [hep-ph]].


\bibitem{Guo:2012tv}
F.~K.~Guo and U.~G.~Meissner,
Phys. Rev. D \textbf{86}, 091501 (2012)
doi:10.1103/PhysRevD.86.091501
[arXiv:1208.1134 [hep-ph]].

\bibitem{Olsen:2014maa}
S.~L.~Olsen,
Phys. Rev. D \textbf{91}, no.5, 057501 (2015)
doi:10.1103/PhysRevD.91.057501
[arXiv:1410.6534 [hep-ex]].

\bibitem{Zhou:2015uva}
Z.~Y.~Zhou, Z.~Xiao and H.~Q.~Zhou,
Phys. Rev. Lett. \textbf{115}, no.2, 022001 (2015)
doi:10.1103/PhysRevLett.115.022001
[arXiv:1501.00879 [hep-ph]].

\bibitem{Yu:2017bsj}
G.~L.~Yu, Z.~G.~Wang and Z.~Y.~Li,
Chin. Phys. C \textbf{42}, 4 (2018)
doi:10.1088/1674-1137/42/4/043107
[arXiv:1704.06763 [hep-ph]].

\bibitem{Gross:2022hyw}
F.~Gross, E.~Klempt, S.~J.~Brodsky, A.~J.~Buras, V.~D.~Burkert, G.~Heinrich, K.~Jakobs, C.~A.~Meyer, K.~Orginos and M.~Strickland, \textit{et al.}
[arXiv:2212.11107 [hep-ph]].

\bibitem{Henriques:1976jd}
A.~B.~Henriques, B.~H.~Kellett and R.~G.~Moorhouse,
Phys. Lett. B \textbf{64}, 85-92 (1976)
doi:10.1016/0370-2693(76)90364-6

\bibitem{Konigsmann:1986wc}
K.~Konigsmann,
Phys. Rept. \textbf{139}, 243 (1986)
doi:10.1016/0370-1573(86)90060-8

\bibitem{Lakhina:2006vg}
O.~Lakhina and E.~S.~Swanson,
Phys. Rev. D \textbf{74}, 014012 (2006)
doi:10.1103/PhysRevD.74.014012
[arXiv:hep-ph/0603164 [hep-ph]].

\bibitem{Eichten:2007qx}
E.~Eichten, S.~Godfrey, H.~Mahlke and J.~L.~Rosner,
Rev. Mod. Phys. \textbf{80}, 1161-1193 (2008)
doi:10.1103/RevModPhys.80.1161
[arXiv:hep-ph/0701208 [hep-ph]].

\bibitem{Zhao:2013jza}
C.~W.~Zhao, G.~Li, X.~H.~Liu and F.~L.~Shao,
Eur. Phys. J. C \textbf{73}, 2482 (2013)
doi:10.1140/epjc/s10052-013-2482-y

\bibitem{Guo:2014zva}
P.~Guo, T.~Y\'epez-Mart\'\i{}nez and A.~P.~Szczepaniak,
Phys. Rev. D \textbf{89}, no.11, 116005 (2014)
doi:10.1103/PhysRevD.89.116005
[arXiv:1402.5863 [hep-ph]].

\bibitem{Deng:2016stx}
W.~J.~Deng, H.~Liu, L.~C.~Gui and X.~H.~Zhong,
Phys. Rev. D \textbf{95}, no.3, 034026 (2017)
doi:10.1103/PhysRevD.95.034026
[arXiv:1608.00287 [hep-ph]].

\bibitem{Belle-II:2018jsg}
E.~Kou \textit{et al.} [Belle-II],
PTEP \textbf{2019}, no.12, 123C01 (2019)
[erratum: PTEP \textbf{2020}, no.2, 029201 (2020)]
doi:10.1093/ptep/ptz106
[arXiv:1808.10567 [hep-ex]].

\bibitem{Hoferichter:2020lap}
M.~Hoferichter and P.~Stoffer,
JHEP \textbf{05}, 159 (2020)
doi:10.1007/JHEP05(2020)159
[arXiv:2004.06127 [hep-ph]].

\bibitem{Yao:2021lus}
X.~Yao,
Int. J. Mod. Phys. A \textbf{36}, no.20, 2130010 (2021)
doi:10.1142/S0217751X21300106
[arXiv:2102.01736 [hep-ph]].

\bibitem{Ganbold:2021nvj}
G.~Ganbold, T.~Gutsche, M.~A.~Ivanov and V.~E.~Lyubovitskij,
Phys. Rev. D \textbf{104}, no.9, 094048 (2021)
doi:10.1103/PhysRevD.104.094048
[arXiv:2107.08774 [hep-ph]].

\bibitem{Hong:2022sht}
K.~H.~Hong, H.~C.~Kim and U.~Yakhshiev,
PTEP \textbf{2022}, no.10, 103D02 (2022)
doi:10.1093/ptep/ptac131
[arXiv:2208.01851 [hep-ph]].

\bibitem{Babiarz:2019sfa}
I.~Babiarz, V.~P.~Goncalves, R.~Pasechnik, W.~Sch\"afer and A.~Szczurek,
Phys. Rev. D \textbf{100}, no.5, 054018 (2019)
doi:10.1103/PhysRevD.100.054018
[arXiv:1908.07802 [hep-ph]].

\bibitem{Feng:2015uha}
F.~Feng, Y.~Jia and W.~L.~Sang,
Phys. Rev. Lett. \textbf{115}, no.22, 222001 (2015)
doi:10.1103/PhysRevLett.115.222001
[arXiv:1505.02665 [hep-ph]].

\bibitem{Feng:2017hlu}
F.~Feng, Y.~Jia and W.~L.~Sang,
Phys. Rev. Lett. \textbf{119}, no.25, 252001 (2017)
doi:10.1103/PhysRevLett.119.252001
[arXiv:1707.05758 [hep-ph]].

\bibitem{Abreu:2022cco}
S.~Abreu, M.~Becchetti, C.~Duhr and M.~A.~Ozcelik,
JHEP \textbf{02}, 250 (2023)
doi:10.1007/JHEP02(2023)250
[arXiv:2211.08838 [hep-ph]].

\bibitem{Chen:2016bpj}
J.~Chen, M.~Ding, L.~Chang and Y.~x.~Liu,
Phys. Rev. D \textbf{95}, no.1, 016010 (2017)
doi:10.1103/PhysRevD.95.016010
[arXiv:1611.05960 [nucl-th]].

\bibitem{Li:2021ejv}
Y.~Li, M.~Li and J.~P.~Vary,
Phys. Rev. D \textbf{105}, no.7, L071901 (2022)
doi:10.1103/PhysRevD.105.L071901
[arXiv:2111.14178 [hep-ph]].

\bibitem{Leitao:2017mlx}
S.~Leit\~ao, A.~Stadler, M.~T.~Pe\~na and E.~P.~Biernat,
Phys. Rev. D \textbf{96}, no.7, 074007 (2017)
doi:10.1103/PhysRevD.96.074007
[arXiv:1707.09303 [hep-ph]].

\bibitem{Fischer:2014cfa}
C.~S.~Fischer, S.~Kubrak and R.~Williams,
Eur. Phys. J. A \textbf{51}, 10 (2015)
doi:10.1140/epja/i2015-15010-7
[arXiv:1409.5076 [hep-ph]].

\bibitem{Li:2018uif}
M.~Li, Y.~Li, P.~Maris and J.~P.~Vary,
Phys. Rev. D \textbf{98}, no.3, 034024 (2018)
doi:10.1103/PhysRevD.98.034024
[arXiv:1803.11519 [hep-ph]].

\bibitem{Li:2019data}
Li, Yang (2019), “Heavy quarkonium light front wave functions from basis light-front quantization with a running coupling”, Mendeley Data, V2, doi: 10.17632/cjs4ykv8cv.2

\bibitem{Maris:2020wew}
P.~Maris, S.~Jia, M.~Li, Y.~Li, S.~Tang and J.~P.~Vary,
PoS \textbf{LC2019}, 007 (2020)
doi:10.22323/1.374.0007
[arXiv:2002.06489 [nucl-th]].

\bibitem{Vary:2009gt}
J.~P.~Vary, H.~Honkanen, J.~Li, P.~Maris, S.~J.~Brodsky, A.~Harindranath, G.~F.~de Teramond, P.~Sternberg, E.~G.~Ng and C.~Yang,
Phys. Rev. C \textbf{81}, 035205 (2010)
doi:10.1103/PhysRevC.81.035205
[arXiv:0905.1411 [nucl-th]].

\bibitem{Li:2017mlw}
Y.~Li, P.~Maris and J.~P.~Vary,
Phys. Rev. D \textbf{96}, 016022 (2017)
doi:10.1103/PhysRevD.96.016022
[arXiv:1704.06968 [hep-ph]].

\bibitem{Adhikari:2018umb}
L.~Adhikari, Y.~Li, M.~Li and J.~P.~Vary,
Phys. Rev. C \textbf{99}, no.3, 035208 (2019)
doi:10.1103/PhysRevC.99.035208
[arXiv:1809.06475 [hep-ph]].

\bibitem{Chen:2018vdw}
G.~Chen, Y.~Li, K.~Tuchin and J.~P.~Vary,
Phys. Rev. C \textbf{100}, no.2, 025208 (2019)
doi:10.1103/PhysRevC.100.025208
[arXiv:1811.01782 [nucl-th]].

\bibitem{Lan:2019img}
J.~Lan, C.~Mondal, M.~Li, Y.~Li, S.~Tang, X.~Zhao and J.~P.~Vary,
Phys. Rev. D \textbf{102}, no.1, 014020 (2020)
doi:10.1103/PhysRevD.102.014020
[arXiv:1911.11676 [nucl-th]].

\bibitem{Lappi:2020ufv}
T.~Lappi, H.~M\"antysaari and J.~Penttala,
Phys. Rev. D \textbf{102}, no.5, 054020 (2020)
doi:10.1103/PhysRevD.102.054020
[arXiv:2006.02830 [hep-ph]].

\bibitem{Babiarz:2023ebe}
I.~Babiarz, R.~Pasechnik, W.~Sch\"afer and A.~Szczurek,
Phys. Rev. D \textbf{107}, no.7, L071503 (2023)
doi:10.1103/PhysRevD.107.L071503
[arXiv:2303.09175 [hep-ph]].




\bibitem{ParticleDataGroup:2022pth}
R.~L.~Workman \textit{et al.} [Particle Data Group],
PTEP \textbf{2022}, 083C01 (2022)
doi:10.1093/ptep/ptac097

\bibitem{Belle:2021nuv}
X.~L.~Wang \textit{et al.} [Belle],
Phys. Rev. D \textbf{105}, no.11, 112011 (2022)
doi:10.1103/PhysRevD.105.112011
[arXiv:2105.06605 [hep-ex]].

\bibitem{Dudek:2006ej}
J.~J.~Dudek, R.~G.~Edwards and D.~G.~Richards,
Phys. Rev. D \textbf{73}, 074507 (2006)
doi:10.1103/PhysRevD.73.074507
[arXiv:hep-ph/0601137 [hep-ph]].

\bibitem{Drell:1969km}
S.~D.~Drell and T.~M.~Yan,
Phys. Rev. Lett. \textbf{24}, 181-185 (1970)
doi:10.1103/PhysRevLett.24.181

\bibitem{Brodsky:1997de}
S.~J.~Brodsky, H.~C.~Pauli and S.~S.~Pinsky,
Phys. Rept. \textbf{301}, 299-486 (1998)
doi:10.1016/S0370-1573(97)00089-6
[arXiv:hep-ph/9705477 [hep-ph]].

\bibitem{Carbonell:1998rj}
J.~Carbonell, B.~Desplanques, V.~A.~Karmanov and J.~F.~Mathiot,
Phys. Rept. \textbf{300}, 215-347 (1998)
doi:10.1016/S0370-1573(97)00090-2
[arXiv:nucl-th/9804029 [nucl-th]].

\bibitem{Lan:2021wok}
J.~Lan \textit{et al.} [BLFQ],
Phys. Lett. B \textbf{825}, 136890 (2022)
doi:10.1016/j.physletb.2022.136890
[arXiv:2106.04954 [hep-ph]].

\bibitem{CLEO:2005efp}
N.~E.~Adam \textit{et al.} [CLEO],
Phys. Rev. Lett. \textbf{94}, 232002 (2005)
doi:10.1103/PhysRevLett.94.232002
[arXiv:hep-ex/0503028 [hep-ex]].

\bibitem{BESIII:2017gcu}
M.~Ablikim \textit{et al.} [BESIII],
Phys. Rev. D \textbf{96}, no.3, 032001 (2017)
doi:10.1103/PhysRevD.96.032001
[arXiv:1703.00077 [hep-ex]].

\bibitem{Whitaker:1976hb}
J.~S.~Whitaker, W.~M.~Tanenbaum, G.~S.~Abrams, M.~S.~Alam, A.~Boyarski, M.~Breidenbach, W.~Chinowsky, R.~DeVoe, G.~J.~Feldman and C.~E.~Friedberg, \textit{et al.}
Phys. Rev. Lett. \textbf{37}, 1596 (1976)
doi:10.1103/PhysRevLett.37.1596

\bibitem{Biddick:1977sv}
C.~J.~Biddick, T.~H.~Burnett, G.~E.~Masek, E.~S.~Miller, J.~G.~Smith, J.~P.~Stronski, M.~K.~Sullivan, W.~Vernon, D.~H.~Badtke and B.~A.~Barnett, \textit{et al.}
Phys. Rev. Lett. \textbf{38}, 1324 (1977)
doi:10.1103/PhysRevLett.38.1324

\bibitem{Gaiser:1985ix}
J.~Gaiser, E.~D.~Bloom, F.~Bulos, G.~Godfrey, C.~M.~Kiesling, W.~S.~Lockman, M.~Oreglia, D.~L.~Scharre, C.~Edwards and R.~Partridge, \textit{et al.}
Phys. Rev. D \textbf{34}, 711 (1986)
doi:10.1103/PhysRevD.34.711

\bibitem{CLEO:2004cbu}
S.~B.~Athar \textit{et al.} [CLEO],
Phys. Rev. D \textbf{70}, 112002 (2004)
doi:10.1103/PhysRevD.70.112002
[arXiv:hep-ex/0408133 [hep-ex]].



\bibitem{CLEO:2006nor}
R.~A.~Briere \textit{et al.} [CLEO],
Phys. Rev. D \textbf{74}, 031106 (2006)
doi:10.1103/PhysRevD.74.031106
[arXiv:hep-ex/0605070 [hep-ex]].

\bibitem{BESIII:2015cby}
M.~Ablikim \textit{et al.} [BESIII],
Phys. Lett. B \textbf{753}, 103-109 (2016)
doi:10.1016/j.physletb.2015.11.074
[arXiv:1511.01203 [hep-ex]].

\bibitem{BESIII:2012urf}
M.~Ablikim \textit{et al.} [BESIII],
Phys. Rev. D \textbf{86}, 092009 (2012)
doi:10.1103/PhysRevD.86.092009
[arXiv:1209.4963 [hep-ex]].

\bibitem{BESIII:2022tfo}
M.~Ablikim \textit{et al.} [BESIII],
Phys. Rev. D \textbf{106}, no.7, 072007 (2022)
doi:10.1103/PhysRevD.106.072007
[arXiv:2204.09413 [hep-ex]].

\bibitem{CLEO:2008ero}
S.~Dobbs \textit{et al.} [CLEO],
Phys. Rev. Lett. \textbf{101}, 182003 (2008)
doi:10.1103/PhysRevLett.101.182003
[arXiv:0805.4599 [hep-ex]].

\bibitem{BESIII:2010gid}
M.~Ablikim \textit{et al.} [BESIII],
Phys. Rev. Lett. \textbf{104}, 132002 (2010)
doi:10.1103/PhysRevLett.104.132002
[arXiv:1002.0501 [hep-ex]].

\bibitem{CLEO:2005vqq}
J.~L.~Rosner \textit{et al.} [CLEO],
Phys. Rev. Lett. \textbf{95}, 102003 (2005)
doi:10.1103/PhysRevLett.95.102003
[arXiv:hep-ex/0505073 [hep-ex]].

\bibitem{Dudek:2009kk}
J.~J.~Dudek, R.~Edwards and C.~E.~Thomas,
Phys. Rev. D \textbf{79}, 094504 (2009)
doi:10.1103/PhysRevD.79.094504
[arXiv:0902.2241 [hep-ph]].

\bibitem{Chen:2011kpa}
Y.~Chen, D.~C.~Du, B.~Z.~Guo, N.~Li, C.~Liu, H.~Liu, Y.~B.~Liu, J.~P.~Ma, X.~F.~Meng and Z.~Y.~Niu, \textit{et al.}
Phys. Rev. D \textbf{84}, 034503 (2011)
doi:10.1103/PhysRevD.84.034503
[arXiv:1104.2655 [hep-lat]].

\bibitem{Becirevic:2012dc}
D.~Becirevic and F.~Sanfilippo,
JHEP \textbf{01}, 028 (2013)
doi:10.1007/JHEP01(2013)028
[arXiv:1206.1445 [hep-lat]].

\bibitem{Li:2021gze}
N.~Li, C.~C.~Liu and Y.~J.~Wu,
EPL \textbf{133}, no.1, 11001 (2021)
doi:10.1209/0295-5075/133/11001

\bibitem{Delaney:2023fsc}
J.~Delaney, C.~E.~Thomas and S.~M.~Ryan,
[arXiv:2301.08213 [hep-lat]].

\bibitem{Brodsky:2014yha}
S.~J.~Brodsky, G.~F.~de Teramond, H.~G.~Dosch and J.~Erlich,
Phys. Rept. \textbf{584}, 1-105 (2015)
doi:10.1016/j.physrep.2015.05.001
[arXiv:1407.8131 [hep-ph]].

\bibitem{Cisek:2014ala}
A.~Cisek, W.~Sch\"afer and A.~Szczurek,
JHEP \textbf{04}, 159 (2015)
doi:10.1007/JHEP04(2015)159
[arXiv:1405.2253 [hep-ph]].


\bibitem{Chen:2016dlk}
G.~Chen, Y.~Li, P.~Maris, K.~Tuchin and J.~P.~Vary,
Phys. Lett. B \textbf{769}, 477-484 (2017)
doi:10.1016/j.physletb.2017.04.024
[arXiv:1610.04945 [nucl-th]].

\bibitem{Goncalves:2017wgg}
V.~P.~Gon\c{c}alves, M.~V.~T.~Machado, B.~D.~Moreira, F.~S.~Navarra and G.~S.~dos Santos,
Phys. Rev. D \textbf{96}, no.9, 094027 (2017)
doi:10.1103/PhysRevD.96.094027
[arXiv:1710.10070 [hep-ph]].

\bibitem{Aschenauer:2017jsk}
E.~C.~Aschenauer, S.~Fazio, J.~H.~Lee, H.~Mantysaari, B.~S.~Page, B.~Schenke, T.~Ullrich, R.~Venugopalan and P.~Zurita,
Rept. Prog. Phys. \textbf{82}, no.2, 024301 (2019)
doi:10.1088/1361-6633/aaf216
[arXiv:1708.01527 [nucl-ex]].

\bibitem{Goncalves:2020ywm}
V.~P.~Goncalves, D.~E.~Martins and C.~R.~Sena,
Nucl. Phys. A \textbf{1004}, 122055 (2020)
doi:10.1016/j.nuclphysa.2020.122055
[arXiv:2008.03145 [hep-ph]].

\bibitem{Anderle:2021wcy}
D.~P.~Anderle, V.~Bertone, X.~Cao, L.~Chang, N.~Chang, G.~Chen, X.~Chen, Z.~Chen, Z.~Cui and L.~Dai, \textit{et al.}
Front. Phys. (Beijing) \textbf{16}, no.6, 64701 (2021)
doi:10.1007/s11467-021-1062-0
[arXiv:2102.09222 [nucl-ex]].

\bibitem{AbdulKhalek:2021gbh}
R.~Abdul Khalek, A.~Accardi, J.~Adam, D.~Adamiak, W.~Akers, M.~Albaladejo, A.~Al-bataineh, M.~G.~Alexeev, F.~Ameli and P.~Antonioli, \textit{et al.}
Nucl. Phys. A \textbf{1026}, 122447 (2022)
doi:10.1016/j.nuclphysa.2022.122447
[arXiv:2103.05419 [physics.ins-det]].

\bibitem{Mantysaari:2022kdm}
H.~M\"antysaari and J.~Penttala,
JHEP \textbf{08}, 247 (2022)
doi:10.1007/JHEP08(2022)247
[arXiv:2204.14031 [hep-ph]].

\bibitem{Abir:2023fpo}
R.~Abir, I.~Akushevich, T.~Altinoluk, D.~P.~Anderle, F.~P.~Aslan, A.~Bacchetta, B.~Balantekin, J.~Barata, M.~Battaglieri and C.~A.~Bertulani, \textit{et al.}
[arXiv:2305.14572 [hep-ph]].

\bibitem{Accardi:2023chb}
A.~Accardi, P.~Achenbach, D.~Adhikari, A.~Afanasev, C.~S.~Akondi, N.~Akopov, M.~Albaladejo, H.~Albataineh, M.~Albrecht and B.~Almeida-Zamora, \textit{et al.}
[arXiv:2306.09360 [nucl-ex]].

\bibitem{Babiarz:2023cac}
I.~Babiarz, V.~P.~Goncalves, W.~Sch\"afer and A.~Szczurek,
Phys. Lett. B \textbf{843}, 138046 (2023)
doi:10.1016/j.physletb.2023.138046
[arXiv:2306.00754 [hep-ph]].



\end{thebibliography}
\end{document}